\documentclass[sigconf]{acmart}
\AtBeginDocument{%
  }

\acmSubmissionID{103}

\copyrightyear{2024}
\acmYear{2024}
\setcopyright{acmlicensed}
\acmConference[SA Art Papers '24]{SIGGRAPH Asia 2024 Art Papers}{December 3--6, 2024}{Tokyo, Japan}
\acmBooktitle{SIGGRAPH Asia 2024 Art Papers (SA Art Papers '24), December 3--6, 2024, Tokyo, Japan}
\acmDOI{10.1145/3680530.3695433}
\acmISBN{979-8-4007-1133-6/24/12}

\begin{document}

\title{\textit{AI-rays}: Exploring Bias in the Gaze of AI Through a Multimodal Interactive Installation}

\author{Ziyao Gao}
\email{zgao662@connect.hkust-gz.edu.cn}
\orcid{0009-0007-2424-0796}
\affiliation{%
  \institution{The Hong Kong University of Science and Technology (Guangzhou)}
  \city{Guangzhou}
  \country{China}
}

\author{Yiwen Zhang}
\email{yzhang452@connect.hkust-gz.edu.cn}
\orcid{0009-0007-7473-7522}
\affiliation{%
  \institution{The Hong Kong University of Science and Technology (Guangzhou)}
  \city{Guangzhou}
  \country{China}}

\author{Ling Li}
\email{lli297@connect.hkust-gz.edu.cn}
\orcid{0009-0003-6899-1314}
\affiliation{%
  \institution{The Hong Kong University of Science and Technology (Guangzhou)}
  \city{Guangzhou}
  \country{China}
}

\author{Theodoros Papatheodorou}
\email{theodoros@hkust-gz.edu.cn}
\orcid{0000-0003-2515-4893}
\affiliation{%
  \institution{The Hong Kong University of Science and Technology (Guangzhou)}
  \city{Guangzhou}
  \country{China}
}

\author{Wei Zeng}
\email{weizeng@hkust-gz.edu.cn}
\orcid{0000-0002-5600-8824}
\authornote{Wei Zeng is the corresponding author}
\affiliation{%
  \institution{The Hong Kong University of Science and Technology (Guangzhou)}
  \city{Guangzhou}
  \country{China}
}
\affiliation{
  \institution{The Hong Kong University of Science and Technology}
  \city{Hong Kong SAR}
  \country{China}
}

\renewcommand{\shortauthors}{Gao et al.}

\begin{abstract}
Data surveillance has become more covert and pervasive with AI algorithms, which can result in biased social classifications. Appearance offers intuitive identity signals, but what does it mean to let AI observe and speculate on them? We introduce AI-rays, an interactive installation where AI generates speculative identities from participants‘ appearance which are expressed through synthesized personal items placed in participants' bags. It uses speculative X-ray visions to contrast reality with AI-generated assumptions, metaphorically highlighting AI's scrutiny and biases. AI-rays promotes discussions on modern surveillance and the future of human-machine reality through a playful, immersive experience exploring AI biases.
\end{abstract}



\begin{teaserfigure}
  \centering 
  \includegraphics[width=0.98\textwidth]{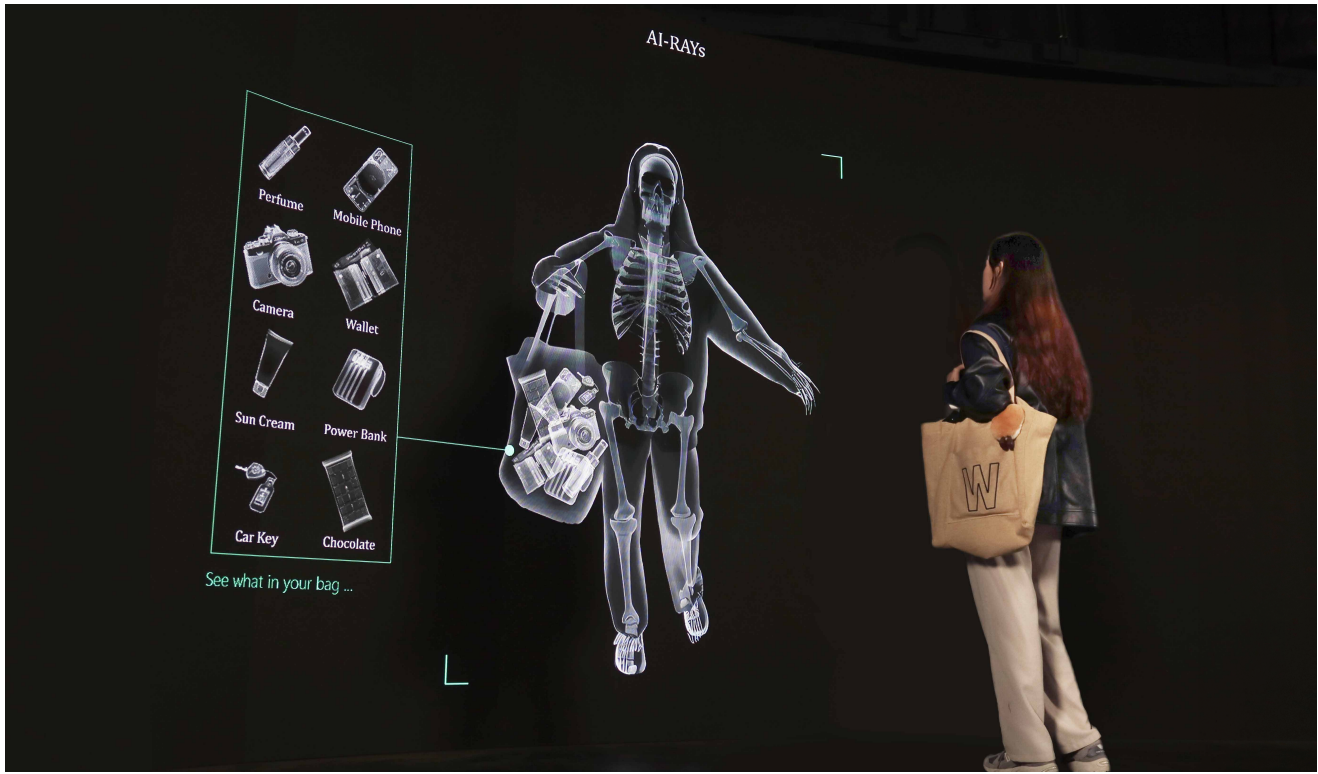}
  \caption{\textbf{AI-rays, the interactive art installation converting participants’ images into speculative generative X-ray images where hypothetical personal items are added into the bags.
  }}
  \Description{AI-rays, the interactive art installation converting participants’ images into speculative generative X-ray images where hypothetical personal items are added into the bags.
  }
  \label{fig:teaser}
\end{teaserfigure}


\maketitle

\section{Inspiration}
Bias is inherent in human society.
In the digital age, ubiquitous data surveillance and the extensive use of algorithms continuously to collect and analyze data about individuals and society, cyclically introduce and reinforce human biases in machine learning models.
Biased processes are deeply embedded in everyday life, arising from various signals such as searching behaviors, click interactions, and, more seriously, physical appearance.
However, the constant assessment for signals is invisible to the public, making it difficult to discern when and how the biases occur, which groups they affect, and the potential cascading consequences they may cause~\cite{ntoutsi2020bias}.
Some artworks externalize these internal processes. 
For example, Kyle McDonald's 
\textit{Facework}~\cite{facework_2020} uses sliders and numbers to visualize how algorithms assess appearance attributes.

Appearance offers natural signals for identifying individuals.
People cultivate their self-perception partly by managing their appearance, while also instinctively making rapid judgments based on others'~\cite{zebrowitz2015faces}. 
The signals can potentially convey various aspects of a person, such as identity, personality, interests, and economic status.

Human understanding of appearance-based signals is grounded in personal experiences and cultural context. However, as exemplified by \textit{Signs that Say What You Want Them To Say and Not Signs that Say What Someone Else Wants You To Say}~\cite{fowler2013once} from Gillian Wearing, it might not represent people’s true selves.
In the context of contemporary digital society, 
what does widespread AI imply when relying on appearance signals and identity labels?

Numerous cases have demonstrated that specific appearance signals can implicitly correlate with biased social sorting, causing injustice.
For example, AI predictive policing overestimates recidivism risk for black people~\cite{predictivePolicing2016}, recruitment engines prefer male candidates for tech jobs~\cite{Amazon2018}, and AI beauty contests favor white winners~\cite{beautyContest2016}.
Nowadays, machine scrutiny is pervasive and constant. 
How do machines interpret our appearance cues? Who is putting that speculation to use? 
Does the meaning of appearance signal change when machines, not humans, observe us?

\section{Concept Design}

\subsection{Identity Narratives of Personal Items}
Since machine bias is typically reflected within groups rather than individuals, its perception and interpretation are often ambiguous. This ambiguity can also manifest in the depiction of identity through personal items.
In his series \textit{Buying Everything on You}~\cite{LiuChuang}, Chuang Liu conveys symbolic profiles of new immigrants in Shenzhen's labor market by showcasing complete sets of their clothes and personal items. These narratives invite audiences to speculate on the identities and lives of the previous owners, fostering non-verbal communication across different eras and cultures.
The connection between items and identities also applies to our work, giving viewers the freedom to interpret.

\subsection{Semantic Transformations of AI-powered X-ray Images}
In the real world, X-rays honestly expose the structure and content of objects. 
Leveraging this characteristic, Nick Veasey practices X-ray photography to challenge society's obsession with external appearances.
His \textit{X-ray Voyeurism}~\cite{NickVoyeurism} series peers into people’s belongings beneath clothing, sparking discussions about the inconsistency between appearances and personal essence. 

Our work also utilizes the unique visual effects of X-rays for defamiliarization, promoting a re-examination of audiences' body image. 
Unlike traditional ones, our generative X-ray images present augmented, speculative X-ray visions.
In the wave of AI-generated art, the nature of images is being altered and debated. For example, Weidi Zhang's \textit{Ray}~\cite{Ray} connects people with AI-powered Rayograph to explore image semantics transformation. 
In our work, the generated X-ray images actually highlight the nature of the data algorithms and AI processes.

Based on these considerations, 
we presented an interactive installation, \textit{\textbf{AI-rays}}. 
Specifically, we enable the machine to observe and analyze audience images from certain perspectives, 
crafting personal items into their bags, and subvert and extend new X-ray images to convey the invasive potential of AI and speculate about the underlying nature of reality. 
\textit{AI-rays} functions as an AI mirror that enables people to re-examine their image and identity when seen through the eyes of large deep-learning models. It helps people identify biases, reflect on the signals they send and ultimately think of the consequences of the proliferation of data surveillance and AI technologies within human society.
\section{Related Work}

\begin{figure*}[t] 
\centering
\includegraphics[width=0.95\textwidth]{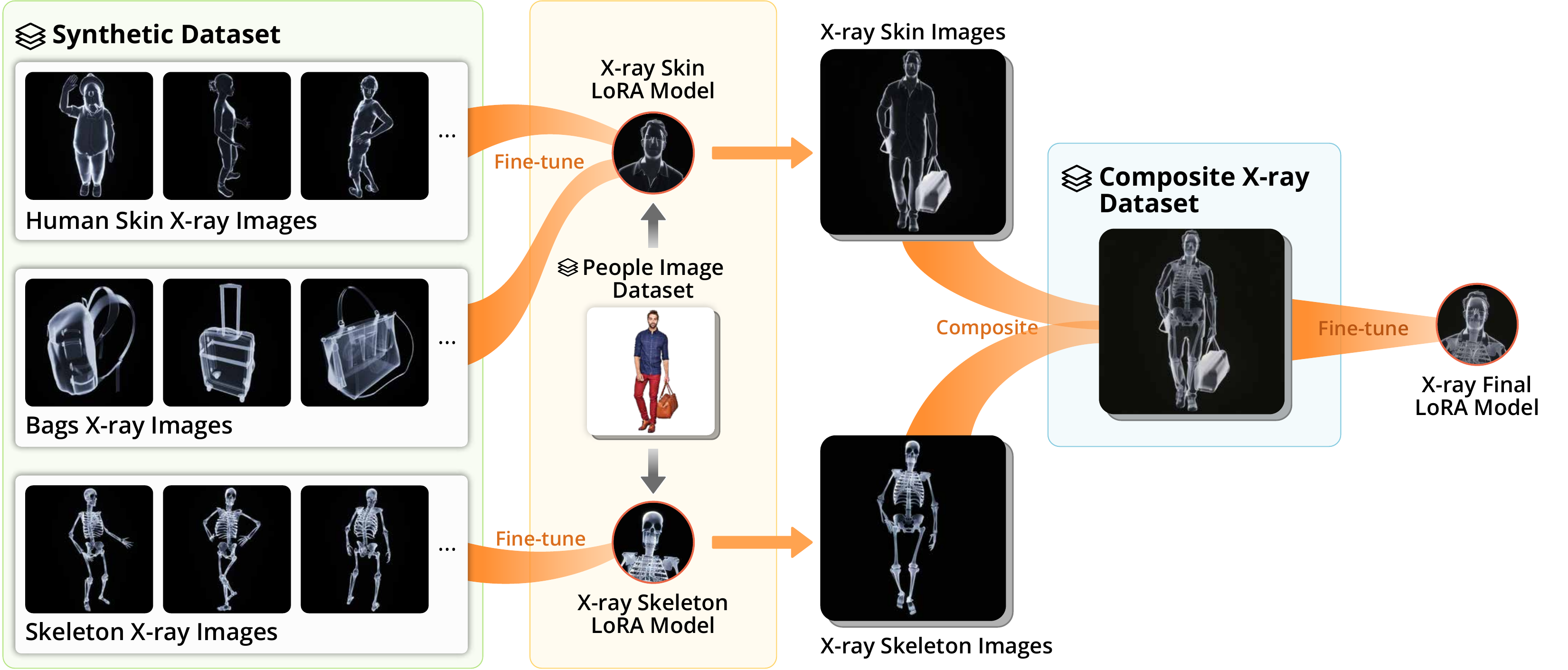}
\vspace{0mm}
\caption{\textbf{The custom pipeline we constructed to build a LoRA model that can output generative X-ray images.}}
\label{fig:sec3-lora}
\end{figure*}

Regarding standards and the power to judge human appearance, 
Mushon Zer-Aviv's interactive installation \textit{The Normalizing Machine}~\cite{mushon2018} 
analyzes public selections of "socially normal images", and gradually aggregates systematic discrimination into the model.
This echoes \textit{Bertillonage}~\cite{ajana2013governing}—the criminal identification system based on anthropometry and photographs invented by police officer Alphonse Bertillon, later adopted by the eugenics movement and Nazis. Today, through the gaze of machines, human appearance is once again scrutinized and classified.

When it comes to the meaning of machine gaze, art experiment \textit{Cheese}~\cite{cheese_2003} from Christian Moeller employs an emotion recognition system to continuously scrutinize actresses' smiles, granting machine the power to associate "smile" with "the performance of sincerity", turning happiness into labor.
In many other instances, machines' understanding of humans is linked to unknown attributes. 
For example, Theodore Watson and Kyle McDonald’s 
\textit{Portrait Machine}~\cite{watson_2014} 
groups audience photos by various visible features to highlight both commonalities and uniqueness.
Mimi Onuoha's \textit{Classification. 01} employs abstract \{\} symbols to represent classification relationships, lighting up when nearby audiences are perceived as similar by unknown standards.
These works explore how machines interpret human data, allowing audiences to glimpse potential algorithmic logic and biases from their own perspectives.

\section{The Installation: AI-rays}

\begin{figure}[h] 
\centering
\includegraphics[width=0.99\columnwidth]{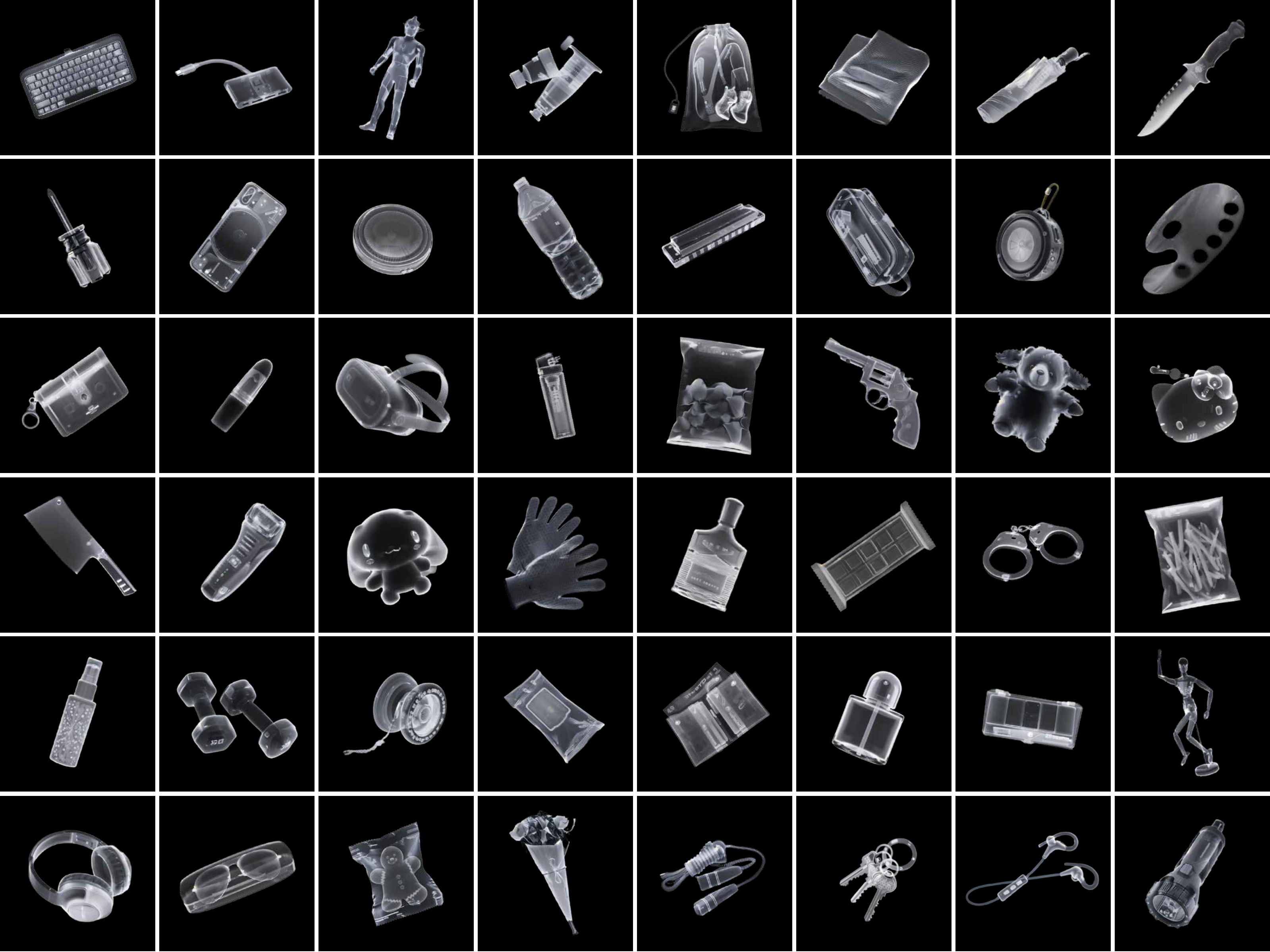}
\vspace{-1mm}
\caption{\textbf{A sample subset from the items dataset used to augment participants’ images.}}
\label{fig:sec3-itemDataset}
\end{figure}

\textit{AI-ray} integrates deep-learning technologies including image understanding, object detection and segmentation, and image generation through specially tuned models. 
Using a custom pipeline, it transforms people into speculative X-ray images. 
Moreover, it visualizes the meaning extracted by machines from people's appearance to keywords and item assignments, then ultimately communicates by crafting personal items onto their bags in the output images.
For example, a bald, muscular man with a beard often gets violent items while a young woman with pigtails is assigned characteristically feminine objects.

\subsection{X-ray Visual Effect Creation}

\begin{figure*}[t] 
\centering
\includegraphics[width=0.98\textwidth]{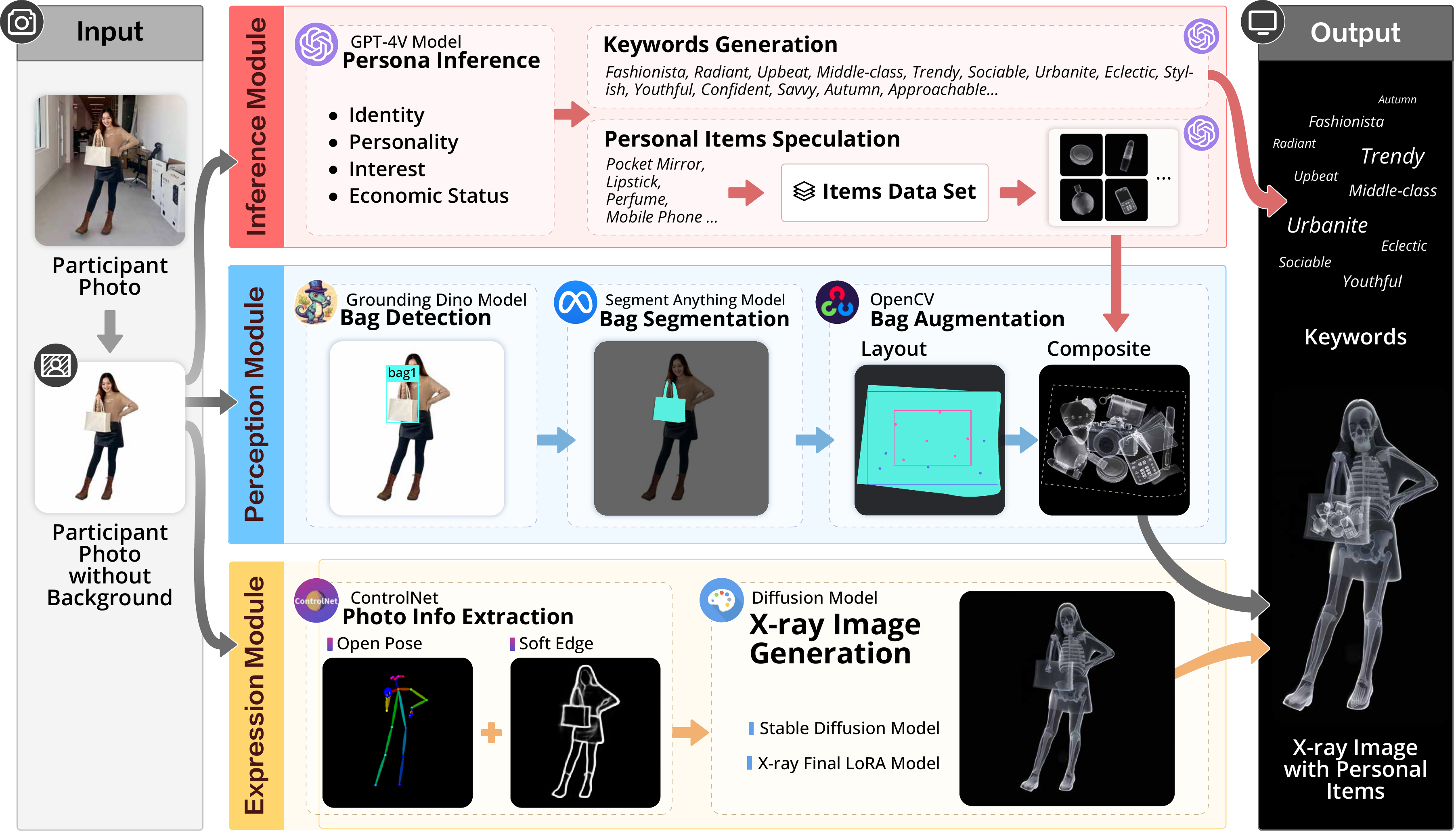}
\vspace{-2mm}
\caption{\textbf{The AI-rays system pipeline accepts a participant’s photo as input (left) and outputs the generative X-ray image along with hypothetical personal items (right). To achieve this it uses three core modules we built that understand, composite, and generate images (center).}}
\label{fig:sec3-system framework}
\end{figure*}

To achieve the X-ray visual effect, we utilize a fine-tuned LoRA model~\cite{hu2021lora} with a diffusion model~\cite{croitoru2023diffusion} to build an artistic X-ray-style and enhance AI’s understanding of human skeletal structure. As shown in Fig.~\ref{fig:sec3-lora}, we first created X-ray-style shaders in Unreal Engine and collected 3D models including human characters, skeletons, and bags, forming a synthetic dataset of 192 rendered X-ray-style images from multiple angles. Using this dataset, we fine-tuned two separate LoRA models, X-ray Skin for human skin and bag images, and X-ray Skeleton for human skeleton images.
We then synthesized 60 full-body shots with bags through both models, getting transparent skin and skeleton images for each. 
By manually compositing these images, we created a new dataset to fine-tune a third model, X-ray Final, which can generate composite skin and skeleton images automatically. 

Next, we generated X-ray versions of various personal items to augment the images. 
We chose pre-generation instead of runtime generation, due to the inherent randomness in Stable Diffusion that can generate some items unidentifiable. 
We constructed a high-quality item dataset containing 148 pre-generated images using the X-ray Final LoRA model and ControlNet~\cite{zhang2023controlNet}.
By allowing GPT-4V~\cite{GPT4V} to iterate through 40 images of people with diverse backgrounds, 
speculating on the objects each person carried, we finally built the list of items. 
A subset of the item images is shown in Fig.~\ref{fig:sec3-itemDataset}.

\subsection{AI-rays Pipeline}

\begin{figure*}[t] 
\centering
\includegraphics[width=0.97\textwidth]{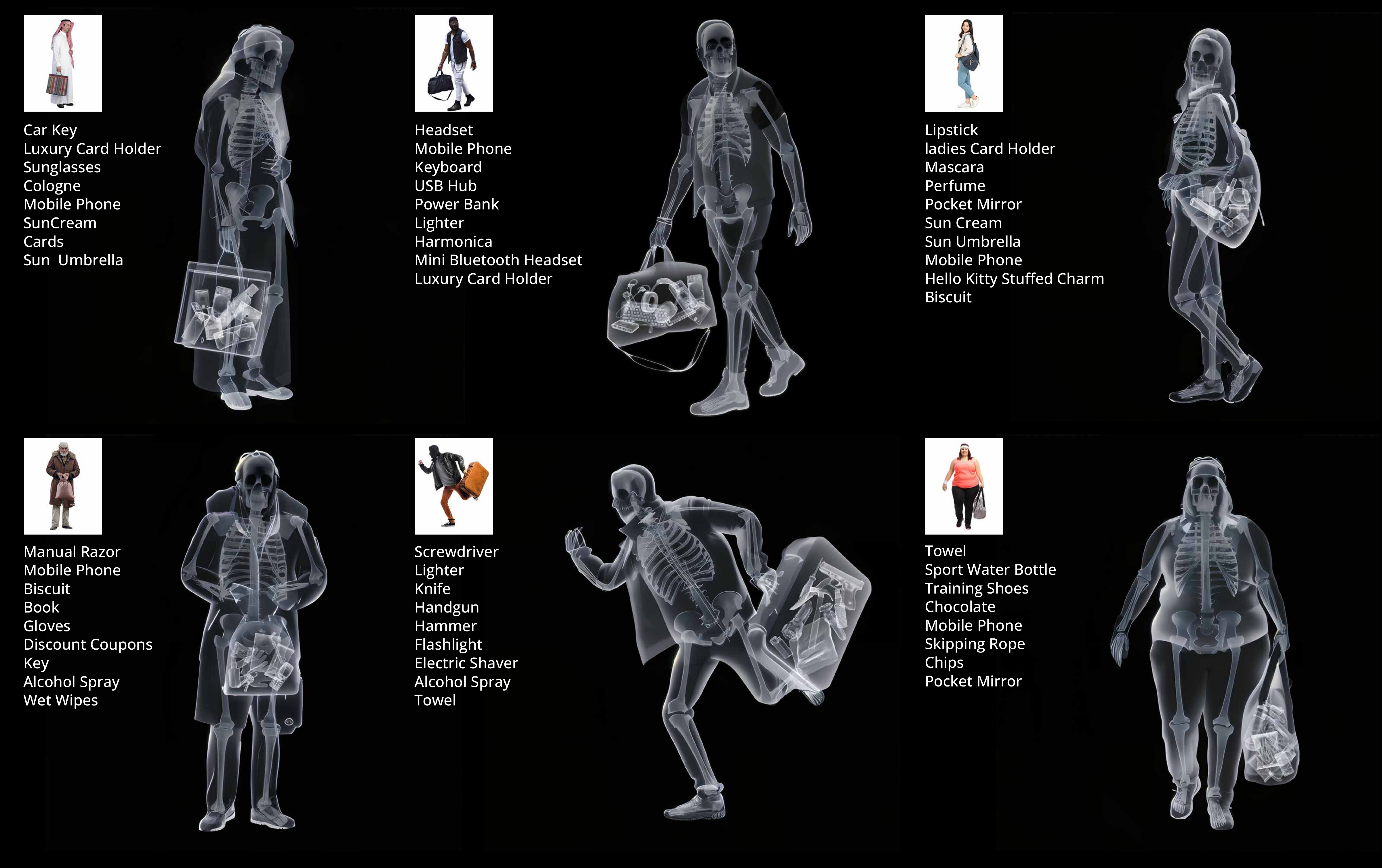}
\vspace{-2mm}
\caption{\textbf{Generated X-ray images with the speculated personal items.} These examples show that women are associated with beauty products and less often with items linked to professions. Additionally, a man in Arab robes is perceived as having a better economic status compared to a man in casual attire.}
\label{fig:sec4-userResult}
\end{figure*}

As shown in Fig.~\ref{fig:sec3-system framework}, \textit{AI-rays} starts by taking the participant’s photo. 
To prevent environmental factors from affecting AI’s understanding, we first remove the photo’s background. Then, submitted it into three parallel modules: \textit{the Inference Module}, \textit{the Perception Module}, and \textit{the Expression Module}.

\begin{itemize}
    \item 
\textit{The Inference Module} is pivotal in understanding human appearance. We utilize GPT-4V to analyze participants on four dimensions: identity, personality, interests, and economic background, constructing a persona for each participant, and then assigning personal items based on it.

\item
\textit{The Perception Module} is responsible for integrating assumed items onto participants’ images. 
Firstly, we employ the Grounding Dino~\cite{liu2023grounding} model to identify the bag. Subsequently, we leverage the Segment Anything Model~\cite{kirillov2023segment} to precisely segment the mask of each bag. 
Finally, using openCV and a custom algorithm we developed, we extract the main region of the mask, calculate layouts, and composite personal items onto bags, ensuring that all items fit well within the bag boundaries and keep the correct relative size to each other.

\item 
\textit{The Expression Module} is tasked with transforming the participant’s photos into X-ray-style using the X-ray Final LoRA model we developed in the earlier stage. 
Initially, we employ the OpenPose and SoftEdge components within ControlNet to extract the posture and outline of the person, serving as a guide for image generation which eliminates the effect of skin and clothes color. 
Then the X-ray-style body image is generated using the X-ray Final LoRA model with the Stable Diffusion model. After upscaling, personal items are then added to that image automatically, and output on a screen.

\end{itemize}

The example output of the AI-rays pipeline is shown in Fig.~\ref{fig:sec4-userResult}.

\subsection{Installation Design}

The installation consists of a camera, a large LED wall screen, and a computer. The LED wall allows participants to immersively engage with their life-size images (Fig.~\ref{fig:teaser}). 

The installation was designed to be intuitive without instructions. 
It attracts participants’ attention with an animated line scan.
When someone approaches, it will be activated.
The generation of the output image takes about ten seconds. 
While the image is being prepared and composited, keywords output associated with the intermediary steps of the algorithmic pipeline pop-up randomly on the screen.
This is to keep the engagement and highlight the “thought process” involved in generating the specific results. 
Once complete, the generated X-ray image is displayed along with speculated personal items.
To present them more clearly, we also enlarge and arrange the items with their names in a floating layout beside the human image.

\section{Feedback and Discussion}

\subsection{Bias Cases within AI Gaze}

\begin{figure}[t] 
\centering
\includegraphics[width=0.9\columnwidth]{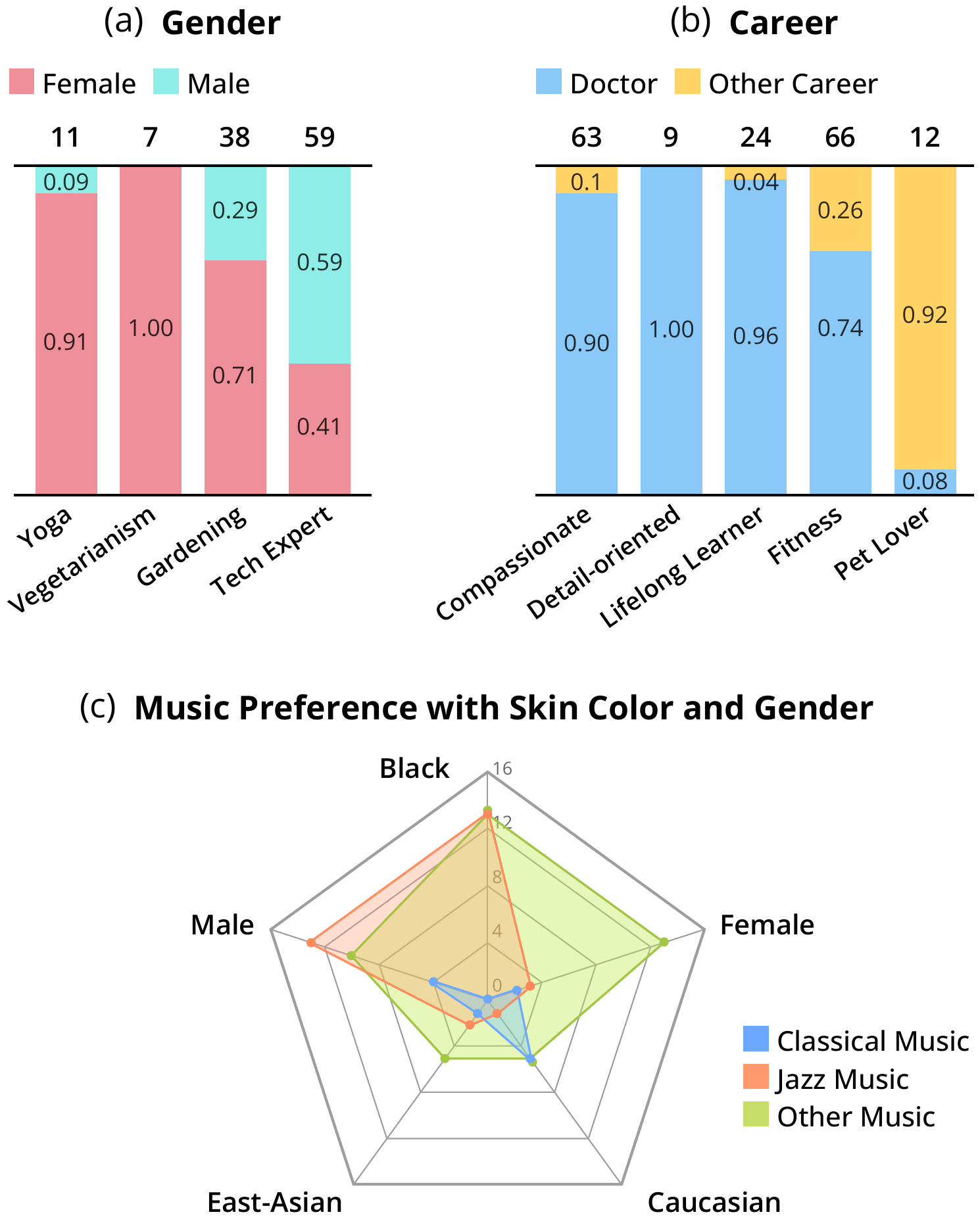}
\vspace{-2mm}
\caption{\textbf{Examples of bias from wider tests based on three representative dimensions: skin color, gender, and occupations using the Inference Module from our pipeline.}}
\label{fig:sec4-inference case}
\end{figure}

To highlight the inherent bias of AI algorithms, we ran the \textit{AI-rays} pipeline on a human dataset of 144 images.
The dataset is diverse along three representative axes: ethnic background, gender, and occupation.
We combined three ethnicities (simply stated: black, caucasian, east Asian), two genders (male and female), and occupations (doctor and non-specific) forming 12 categories (e.g., black male doctor, white female with no specific occupation. etc), and collected 12 portrait photos for each category. 
Through testing, we manually encoded the output keywords, identifying the representative bias cases. 

Fig.~\ref{fig:sec4-inference case}(a) shows the AI tended to associate females with yoga and vegetarianism, potentially mirroring media cultural ideals about female body shapes. 
Noteworthy, its male bias for Tech Expert matches gender biases reported in recruitment algorithms. 
In Fig.~\ref{fig:sec4-inference case}(b), doctors are seen as compassionate, detail-oriented, lifelong learners, and fitness enthusiasts. 
Fig.~\ref{fig:sec4-inference case}(c) shows AI's perceived correlations between musical preferences, skin color, and gender, suggesting females are less interested in music than males, and black people more so than Caucasian and Asian people.
Notably, AI assumes a strong preference for jazz in black males.

\subsection{Human Perceptions of Machine Scrutiny and Bias}

We recruited 15 participants to test our prototype installation and gather feedback.
Participants reported that the experience evoked a sense of scrutiny. 
Lozano-Hemmer's \textit{Surface Tension}~\cite{Surface_1992} explores intrusive detection and control by tracking audiences with a visible electronic eye. In contrast, we created a calmer yet pervasive atmosphere of scrutiny, making individuals "voluntary" subjects of inspection. 
Our experience starts with line scanning, making participants uneasy, "\textit{like entering a rigorous security checkpoint.}" The transparent X-ray images enhance the feeling of being observed. Additionally, larger-than-life skeletal figures alienate the everyday space, one commenting, "\textit{I never imagined standing in front of my skeleton and being looked down upon by it. I felt oppressed.}" 

The intriguing correspondence between personal items and identity introduced a level of interpretative ambiguity, fostering imagination and exploration among participants.
Two females reacted differently to assigned beauty products
—one bemused by stereotypes, stating, "\textit{It assumes I need perfume and lipstick just because I'm a woman, but I never wear makeup.}" The other finding them "\textit{exactly what I use every day.}"
Some desired a second interaction to question if the results were insightful or random. 
Additionally, three participants attempted to disguise themselves from the system's scrutiny. One used a baseball cap and mask, receiving items associated with a suspicious character.
Throughout the processes, participants accepted scrutiny actively or passively, forming personal insights of bias. 
This led to questioning the power of intelligent systems in surveillance and control from the perspective of individual identities and AI’s role.

The experience prompted participants to reflect on their appearance signals and how AI perceived them. 
Most found the inferences relatively accurate and were surprised by AI’s capabilities, 
with some feeling more confident about their self-image. 
However, some worried their personal pursuits of individuality were merely reflections of societal categories. 
A few participants found the results untrue. 
Interestingly, they rated the experience more positively and engaged in discussions afterward.

\textit{Subtitled Public} from Lozano-Hemmer projects critical subtitles onto people, highlighting the violence and asymmetry in observation. 
Similarly, our work introduces continuous, unconsented observation of appearance, but incorporates participants’ understanding of the system’s outputs and self-perception into the experience. 
The ambiguous connection between items and identity avoids overly sharp critiques, transforming classification and judgment into a game. This dynamic process aligns with David Lyon’s theory of surveillance practices in modern society~\cite{lyon2018culture}, which are inherently linked to power but oscillate between care and control without absolute intentions or outcomes.

Apart from revealing algorithmic biases based on appearance signals and discussing the power dynamics of surveillance and bias through a playful approach, this project also explores new possibilities of X-ray vision by creating X-ray images with semantic information. One participant remarked, "\textit{I usually associate X-rays with hospitals and illness. Through this work, I see an alternate reality.}"
Unlike the other art installations that highlight the menacing nature of surveillance in darker tones, our work tries to highlight the AI biases and evoke discussion about surveillance and appearance signals through fun and surprising experiential engagement.
While our approach may be indirect, it ultimately sparks discussions about technical biases, self-image, and the future of human-machine reality.

\begin{acks}
This work was supported by the Guangzhou Basic and Applied Basic Research Foundation (2024A04J6462).
\end{acks}
\balance
\bibliographystyle{ACM-Reference-Format}
\bibliography{Reference}


\end{document}